# Planar nano-optics in anisotropic media: anomalous refraction and diffraction-free lensing of highly confined polaritons

J. Duan[1,2†], G. Álvarez-Pérez[1,2†], A. I. F. Tresguerres-Mata[1], J. Taboada-Gutiérrez[1,2], K. V. Voronin[3], A. Bylinkin[4,5], B. Chang[6], S. Xiao[7], S. Liu[8], J. H. Edgar[8], J. I. Martín[1,2], V. S. Volkov[3,9], R. Hillenbrand[4,10], J. Martín-Sánchez[1,2], A. Y. Nikitin[5,10], P. Alonso-González[1,2]*

[1]*Department of Physics, University of Oviedo, Oviedo 33006, Spain.*
[2]*Center of Research on Nanomaterials and Nanotechnology, CINN (CSIC-Universidad de Oviedo), El Entrego 33940, Spain.*
[3]*Center for Photonics and 2D Materials, Moscow Institute of Physics and Technology, Dolgoprudny 141700, Russia.*
[4]*CIC nanoGUNE BRTA, 20018 Donostia - San Sebastian, Spain.*
[5]*Donostia International Physics Center (DIPC), Donostia/San Sebastián 20018, Spain*
[6]*National Centre for Nano Fabrication and Characterization, Technical University of Denmark, DK-2800 Kgs, Lyngby, Denmark.*
[7]*DTU, Fotonik, Department of Photonics Engineering and Center for Nanostructured Graphene, Technical University of Denmark, DK-2800 Kgs. Lyngby, Denmark.*
[8]*Tim Taylor Department of Chemical Engineering, Kansas State University, Manhattan, KS 66506, USA.*
[9]*GrapheneTek, Skolkovo Innovation Center, Moscow 143026, Russia.*
[10]*IKERBASQUE, Basque Foundation for Science, Bilbao 48013, Spain.*

\* pabloalonso@uniovi.es

† *These authors contributed equally to this work.*

**As one of the most fundamental optical phenomena, refraction between isotropic media is characterized by light bending towards the normal to the boundary when passing from a low- to a high-refractive-index medium. However, in anisotropic media, refraction is a much more exotic phenomenon[1-4]. The most general case of refraction between two anisotropic media remains unexplored, particularly in natural media and at the nanoscale. Here, we visualize and comprehensively study refraction of electromagnetic waves between two strongly anisotropic (hyperbolic) media, and, importantly, we do it with the use of polaritons confined to the nanoscale in a low-loss natural medium, α-MoO$_3$. Our images show refraction of polaritons under the general case in which both the direction of propagation and the wavevector are not collinear. As they traverse planar nanoprisms tailored in α-MoO$_3$, refracted polaritons exhibit non-intuitive directions of propagation, enabling us to unveil an exotic optical effect: bending-free refraction. Furthermore, we succeed in developing the first in-plane refractive hyperlens, which yields foci as small as λ$_p$/6, being λ$_p$ the polariton wavelength (λ$_0$/50 with respect to the wavelength of light in free space). Our results set the grounds for planar nano-optics in strongly anisotropic media, with potential for unprecedented control of the flow of energy at the nanoscale.**

Hyperbolic electromagnetic waves arise as a consequence of the intrinsic anisotropy of the crystal lattice in natural media[1-3] and of the artificially engineered anisotropic dielectric environment in metamaterials[4-12], which leads to a metallic-like response (negative permittivity) along one (two) of the optical axes in such media and a dielectric-like response (positive permittivity) along the other two (one). Despite their fundamental interest and their potential for the development of new optical applications, these exotic waves are still very little explored. In particular, refraction of hyperbolic waves has only been studied for the case in which the incident beam comes from an isotropic medium, typically free space[5-7]. As such, the most general case of refraction, involving hyperbolic waves in which both the incident and the refracted waves exhibit non-collinear wavevector *k* and energy flux *S*, remains experimentally unexplored, particularly at the nanoscale, where the specific case of negative refraction of highly confined polaritons has only recently been theoretically proposed[13-14]. Therefore, the study of the most general case of refraction could extend our capabilities to control the flow of light.

Importantly, the recent discoveries of phonon polaritons (PhPs)[15,16]—light coupled to lattice vibrations— with hyperbolic dispersion in the van der Waals (vdW) crystals h-BN[17-19], α-MoO$_3$[20-25], and α-V$_2$O$_5$[26], have provided unique material platforms to study optical phenomena[27-28] within strongly anisotropic natural media. In particular, PhPs in α-MoO$_3$ feature in-plane hyperbolic propagation, ultra-low losses, and strong confinement, offering the possibility to visualize refraction phenomena directly on the crystal surface and at the nanoscale, which can open new routes in planar nano-optics[29].

The unique properties of polaritons in hyperbolic media can be better understood by analyzing their iso-frequency curve (IFC), a slice of the polariton dispersion surface in momentum-frequency space ($k_x$, $k_y$, $\omega$) by a plane of constant frequency. The IFCs of polaritons in two different hyperbolic media are illustrated in Fig.1a-b. For convenience, we consider these two media defined by the same hyperbolic slab (with representative permittivity $\varepsilon_x$ = -5, $\varepsilon_y$ = 1, $\varepsilon_z$ = 5, see Methods) placed on two different dielectric substrates (with permittivities $\varepsilon_{sub}$ = 1 and $\varepsilon_{sub}$ = 5). In both cases, the IFCs describe open hyperbolas (black and gray curves, respectively). As a result, not all wavevectors ***k*** are allowed in these media, which implies that polaritons cannot propagate along all in-plane directions in real space. In fact, propagation is only allowed within the sectors $|\tan(k_x/k_y)| < \sqrt{-\varepsilon_y/\varepsilon_x}$ limited by the asymptotes of the hyperbola in the ($k_x$, $k_y$) plane (see Fig. 1a-b). Additionally, the Poynting vector ***S*** —which determines the propagation direction of the polariton and is normal to the IFC[30-31]— is not in general collinear with ***k***, as indicated in Fig. 1a (they are collinear only along the x-axis in Fig. 1a). As such, the properties of propagating polaritons in hyperbolic media are very different to those in isotropic media, where the IFCs describe circumferences (see dashed cyan curve in Fig. 1a) and polaritons, as is well-known, are allowed to propagate along all in-plane directions in real space with the same absolute value of the wavevector, ***k***, which is always collinear to ***S***. Importantly, such properties of polaritons propagating in hyperbolic media have a dramatic effect when they refract at a boundary between two different hyperbolic media. Particularly, momentum conservation at the boundary implies that the projection ***k***$_\parallel$ (orange arrows in Figs. 1a-b) of the incident and refracted wavevectors (***k***$_{\text{in}}$ and ***k***$_{\text{out}}$, respectively), must be conserved, giving rise to the generalized Snell's law[32]:

$$k_{\text{in}} \cdot \sin(\theta_{\text{in}-k}) = k_{\text{out}} \cdot \sin(\theta_{\text{out}-k}), \quad (1)$$

where $\theta_{in-k}$ and $\theta_{out-k}$ are the angles that $\boldsymbol{k_{in}}$ and $\boldsymbol{k_{out}}$ form with the normal to the boundary, respectively (see Fig. 1c-d). The propagation directions of the incident and refracted waves are then given by $\boldsymbol{S_{in}}$ and $\boldsymbol{S_{out}}$, respectively, i.e. the directions normal to the hyperbolic IFCs for each case, which are in general non-collinear with $\boldsymbol{k_{in}}$ and $\boldsymbol{k_{out}}$ (Fig. 1b-d), and thus refraction can occur at angles $\theta_{out-S}$ that can be very different from $\theta_{out-k}$. This behavior is in stark contrast to that in isotropic media in which momentum conservation at the boundary implies that refraction occurs always at $\theta_{out-k}= \theta_{out-S}$.

A particular case of anomalous refraction between hyperbolic media for an incident wave with collinear $\boldsymbol{k_{in}}$ and $\boldsymbol{S_{in}}$ is shown in Fig. 1a. We observe that the refracted wave ($\boldsymbol{S_{out}}$) bends away from the direction of $\boldsymbol{S_{in}}$ towards the boundary (Fig. 1c), in contrast to what is expected in isotropic media for a wave passing from a low refractive index to a high refractive index, where $\boldsymbol{S_{out}}$ bends towards the normal to the boundary (see Fig. S8 in the Supplementary Information). Also, the modulus of $\boldsymbol{k_{out}}$ is much larger than that of $\boldsymbol{k_{in}}$, showing a strong wavelength reduction.

The most general case of refraction occurs when $\boldsymbol{k_{in}}$ and $\boldsymbol{S_{in}}$ of the incident wave are not collinear. This case, which has not been tackled experimentally to date, is sketched in Fig. 1d, where a wave impinges at a tilting angle $\theta_{in-S}$ on the boundary between two hyperbolic media. The boundary is also tilted a given angle with respect to the crystal axes. In this case, the refracted wave propagates almost parallelly to the incident wave (i.e. $\theta_{in-S} \approx \theta_{out-S}$), as if the incident wave had been directly transmitted without any change in its propagation direction (black and blue arrows in Fig. 1d). This feature opens the door to bending-free refraction in anisotropic media, which in isotropic media is only possible at normal incidence. In addition, the modulus of $\boldsymbol{k_{out}}$ is much larger than the modulus of $\boldsymbol{k_{in}}$, revealing a strong wavelength reduction due to the hyperbolic shape of the IFCs.

Now, we experimentally demonstrate and comprehensively study the characteristics of anomalous refraction of polaritons propagating in hyperbolic media. This provides the first demonstration of these effects at the nanoscale and/or in a natural medium. To do so, we design and fabricate planar prisms (Methods) in a slab of the naturally hyperbolic van der Waals crystal α-MoO$_3$. We then visualize the propagation of hyperbolic phonon polaritons (HPhPs) passing through them by polariton wavefront mapping using a scattering-type scanning near-field optical microscope (s-SNOM, see Methods). To define the prims, we etch away triangular regions in a silica (SiO$_2$) substrate on top of which we place a 160-nm-thick α-MoO$_3$ slab, thus forming a region α-MoO$_3$/air with a different refractive index, and thus different polaritonic dispersion, with respect to the region α-MoO$_3$/SiO$_2$. The different polaritonic dispersions in these two regions are clearly corroborated by the near-field image of Fig. 2a (left panel), taken at an incident wavelength $\lambda_0 =$ 11.3 μm. Specifically, we observe HPhPs launched inside the prism (highlighted by white dashed lines) by the flake edge (see Methods and Fig.S4 in Supplementary Information), which propagate with collinear $\boldsymbol{k_{in}}$ and $\boldsymbol{S_{in}}$ and longer wavelength $\lambda_{in}$ (white arrow), i.e. smaller wavevector $|k_{in}|$ = $2\pi/\lambda_{in}$ (red arrow), than outside the prism ($\lambda_p$, black arrow), thus revealing the lower refractive index of the prisms. Excitingly, we also observe in the same near-field image that when the HPhPs reach a boundary of the prism tilted at an angle $\theta_{in}\sim 55°$, they refract into the α-MoO$_3$/SiO$_2$ region (note that reflection is expected to be very weak, see Section S8 in Supplementary Information) along a different direction ($\boldsymbol{S_{out-exp}}$, blue arrows) with respect to which their wavefronts are tilted ($\boldsymbol{k_{out-exp}}$, green arrows). Thus, the refracted energy flux and wavevector are not collinear, in agreement with our predictions for hyperbolic refraction in Fig.1a. Importantly, while the wavevector, $k_{out-exp}$, refracts towards the normal (note that polariton launching by the prism boundary can be ruled out, see Supplementary Information), the energy flux, $S_{out-exp}$, bends away from it. This is in stark contrast to what is expected for a wave propagating from a lower refractive index region to a higher refractive index region in isotropic media. In addition, the modulus of the refracted wavevector $k_{out-exp}$ (~ 6.48 µm$^{-1}$) is considerably larger than that of the incident wavevector $k_{in}$ (~2.09 µm$^{-1}$) and that of the polariton wavevector along the x direction outside the prism $k_p$ (~4.08 µm$^{-1}$), revealing a strong wavelength reduction resulting from refraction in hyperbolic media. These findings are in perfect agreement with the case predicted in Figs. 1a and 1c for hyperbolic refraction considering collinear $k_{in}$ and $S_{in}$.

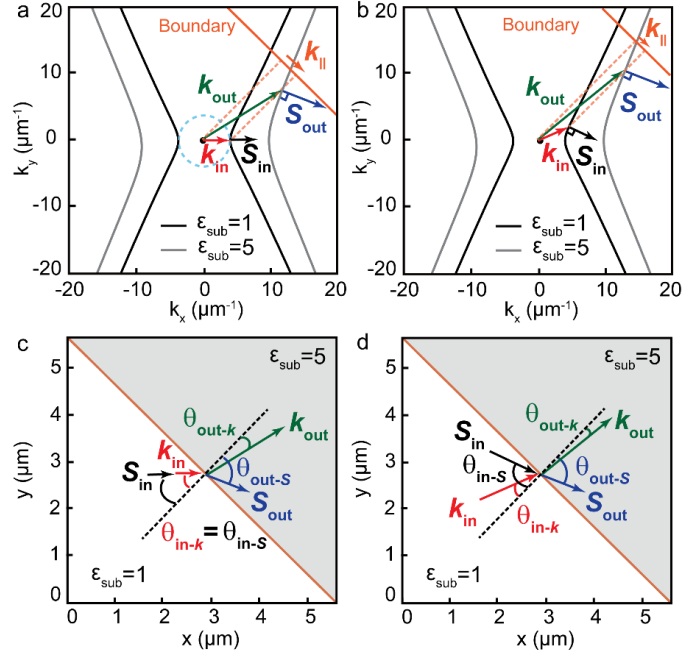

**Fig. 1 | Schematics of refraction of polaritons between two hyperbolic media. a**, IFCs of polaritons propagating in a hyperbolic slab (with $\varepsilon_x = -5$; $\varepsilon_y = 1$; $\varepsilon_z = 5$) placed on two different semi-infinite substrates with $\varepsilon_{sub} = 1$ (black curve) and $\varepsilon_{sub} = 5$ (gray curve) that define two different hyperbolic media (medium 1 and 2, respectively). The incident wave in medium 1 is characterized by collinear $k_{in}$ and $S_{in}$ (as in an isotropic medium, indicated by a dashed cyan circle). Upon refraction into medium 2, momentum conservation at the boundary (orange line), $k_{||}$, is fulfilled by non-collinear $k_{out}$ and $S_{out}$. **b**, The most general case of refraction between two hyperbolic media is represented by an incident wave from medium 1 with non-collinear $k_{in}$ and $S_{in}$ (normal to the IFC). When the wave refracts into medium 2, momentum conservation at the boundary (orange line) is fulfilled by non-collinear $k_{out}$ and $S_{out}$. **c**, Real space illustration of refraction between two hyperbolic media shown in (a) where the incident wave exhibits collinear $k_{in}$ and $S_{in}$, i.e. $\theta_{in-k} = \theta_{in-S}$, giving rise to non-collinear $k_{out}$ and $S_{out}$, i.e. $\theta_{out-S} \neq \theta_{out-k}$. **d**, Real space illustration of the general case of refraction between two hyperbolic media shown in (b) where both the incident and the outgoing wave exhibits non-collinear $k$ and $S$, i.e. $\theta_{in-k} \neq \theta_{in-S}$ and $\theta_{out-k} \neq \theta_{out-S}$. Hyperbolicity gives rise to bending-free refraction, i.e. $\theta_{in-S} \approx \theta_{out-S}$.

To unambiguously verify the anomalous features of refraction between hyperbolic media, we carry out full-wave numerical simulations which mimic our experiments[33] (see Methods). The resulting spatial distribution of the out-of-plane component of the electric field, $\text{Re}(E_z(x,y))$, is plotted in Fig. 2a (middle panel), clearly showing anomalous refraction of both the energy flux ($S_{out}$) and the wavevector ($k_{out}$), in excellent qualitative and quantitative agreement ($|k_{out}|$ ~6.35μm$^{-1}$) with the experimental image in Fig. 2a (left panel). In addition to numerical simulations, we also validate our experimental results by performing analytical calculations[34] (Fig. 2a, right panel) analogous to those shown in Fig. 1. Namely, we calculate the IFCs of HPhPs in α-MoO$_3$/air (gray curve), and α-MoO$_3$/SiO$_2$ (black curve) regions at $\lambda_0$ = 11.3 μm, and applying the condition of momentum (wavevector) conservation at the boundary (orange line), extract the Poynting vector and wavevector of the refracted polaritons. Again, we observe anomalous refraction of the energy flux ($S_{out}$, blue arrow) with a tilted wavevector ($k_{out}$, green arrow) in excellent agreement with the experiment, as well as with the full-wave numerical simulations.

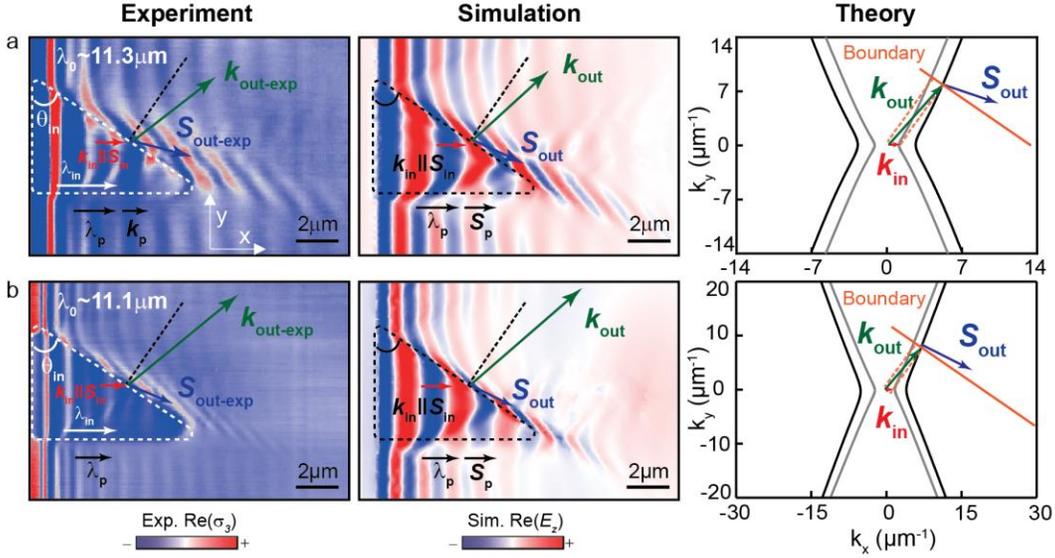

**Fig. 2 | Real-space visualization of refraction between two anisotropic media using highly confined polaritons with collinear incident $k$ and $S$. a**, Experimental Re($\sigma_3(x,y)$) (left column) and simulated Re($E_z(x,y)$) (center column) near-field images of HPhPs propagating in a 160-nm-thick α-MoO$_3$ flake at $\lambda_0$ = 11.3 μm. The white dashed line marks a triangular prism fabricated by etching an air cavity on the SiO$_2$ substrate below the α-MoO$_3$ flake. Upon refraction at a boundary of the prism with an angle $\theta_{in}$ ~ 55°, HPhPs bend away from the normal, $S_{out-exp}$ (blue arrow), with a tilted wavevector $k_{out-exp}$ (green arrow). Compared to non-refracted HPhPs, indicated by $\lambda_p$, the refracted HPhPs are stronger confined (with a wavelength about 1.6 times shorter. From analytic IFCs (right column) of α-MoO$_3$/SiO$_2$ (black hyperbolas) and α-MoO$_3$/air (grey hyperbolas) effective media, and considering momentum conservation at the boundary (orange line), the extracted wavevector and direction of the refracted polaritons, $k_{out}$ and $S_{out}$, respectively, are in good agreement with both experiment and simulation. **b**, Same as (a) for $\lambda_0$ = 11.1 μm. The refracted HPhPs propagate almost parallel to the boundary with a wavelength 2.1 times smaller than $\lambda_p$.

To further analyze anomalous refraction in hyperbolic media, we also perform experiments at a different illuminating wavelength $\lambda_0$ (different tilting angles of the prism boundary $\theta$ are also shown in Fig.S2), as shown in Fig. 2b for $\lambda_0$ = 11.1 μm. Interestingly, we observe that, in this case, both the angular separation between $k_{out-exp}$ and $S_{out-exp}$ and the confinement effect are larger, being the refracted wave ($S_{out-exp}$) almost parallel to the prism boundary and the modulus of $k_{out-exp}$ (12.47 µm⁻¹) about 4 times larger than that of $k_{in}$ (~ 3.14 µm⁻¹) and 2.1 times larger than that of $k_p$ (~ 5.92 µm⁻¹).

Altogether, these results demonstrate the efficient refractive nature of our planar prisms, enabling us to visualize in real-space three important features of highly confined polaritons refracted at the boundary between two hyperbolic media: *i*) large tilting of their wavefronts (given by $k_{out-exp}$) with respect to their propagation direction (given by $S_{out-exp}$), *ii*) counter-intuitive directions of propagation and, *iii*) sub-wavelength confinement (with respect to polaritons along the crystal axes in the same medium).

Such unique features of refracted polaritons in naturally in-plane hyperbolic media opens the door for planar focusing of ultra-confined polaritons. To demonstrate this possibility, we design and fabricate a planar lens in α-MoO₃ (Fig. 3a). As noted above (Fig. 2), when HPhPs with collinear $k_{in}$ and $S_{in}$ (red arrow in Fig. 3b) refract at a boundary with another hyperbolic medium with higher refractive index (such as when passing from α-MoO₃/air to α-MoO₃/SiO₂), they bend away from the normal to the boundary (blue arrow in Fig. 3b). This means that, in the case of considering a prism of triangular shape, as that shown in Fig. 3a, all refracted polaritons (blue arrows) can converge into a single spot, and thus the prism acts as a focusing lens for highly confined polaritons (for the analytical description of general shapes of lenses focusing hyperbolic waves see Supplementary Information). More importantly, as such a lens is based on refraction of HPhPs, the refracted waves can potentially feature infinitely large wavevectors when the boundary is perpendicular to the asymptote of the hyperbolic IFC, which would yield deeply sub-diffractional foci sizes. However, as HPhPs decay exponentially, the intensity at a distant focus can be weak due to propagation losses, which would be more notable for large wavevectors approaching the asymptote of the hyperbolic IFC. Consequently, we designed our triangular lens looking for a compromise between a large refracted wavevector and a long propagation length of the refracted polaritons. According to our theoretical calculations (Fig. S7), this compromise is obtained for an angle of the in-plane wavevector of about 62°. Therefore, we fabricate a triangular prism with boundaries perpendicular to this angle (Fig. 3a). The experimental and simulated near-field images for this lens design upon IR illumination at 11.16 µm are shown in Fig. 3c and 3d, respectively. In both images, we observe the refraction of incident HPhPs with collinear $k_{in}$ and $S_{in}$ (red arrows) at the lens boundaries (black dashed contour) resulting in HPhPs with non-collinear $k_{out}$ and $S_{out}$ (green and blue arrows, respectively) propagating along directions almost parallel to the boundaries (blue arrows), which eventually converge, resulting in a focus. This result is in stark contrast to that observed in a similar lens based on refraction of highly confined polaritons in an isotropic material, such as h-BN, in which refracted polaritons bend towards the normal (Fig. S8), making all them to diverge (Fig. 3e). Furthermore, the near-field images of our hyperbolic lens reveal that the wavevector $k_{out}$ of the refracted HPhPs is much larger than the wavevector $k_{in}$ of the incident HPhPs (note that in the experimental image there is a contribution of tip-launched HPhPs with $2k_{in}$ wavevectors) and the wavevector $k_p$ of HPhPs propagating along the x-direction on the flake (black arrows), which explains that the focus obtained shows a full width at half maximum (FWHM) of ~240 nm (red dots and grey curve in Fig. 3f, for experimental and simulated line profiles, respectively). This size is much smaller than the polariton wavelength ($\lambda_p$) along the x-direction, or the free-space illumination ($\lambda_0$), namely of ~$\lambda_p$/6, or $\lambda_0$/50. This resolution reveals that a diffraction-limited optical system in hyperbolic media can show focusing much smaller than the incident polaritonic wavelength, which again reflects the anomalous behavior of electromagnetic waves in hyperbolic media. Moreover, this result confirms our planar lens based on refraction of HPhPs as a brand-new nano-optical element that greatly exceeds the focusing resolution of any lens based on refraction of highly confined polaritons in isotropic media[35-38].

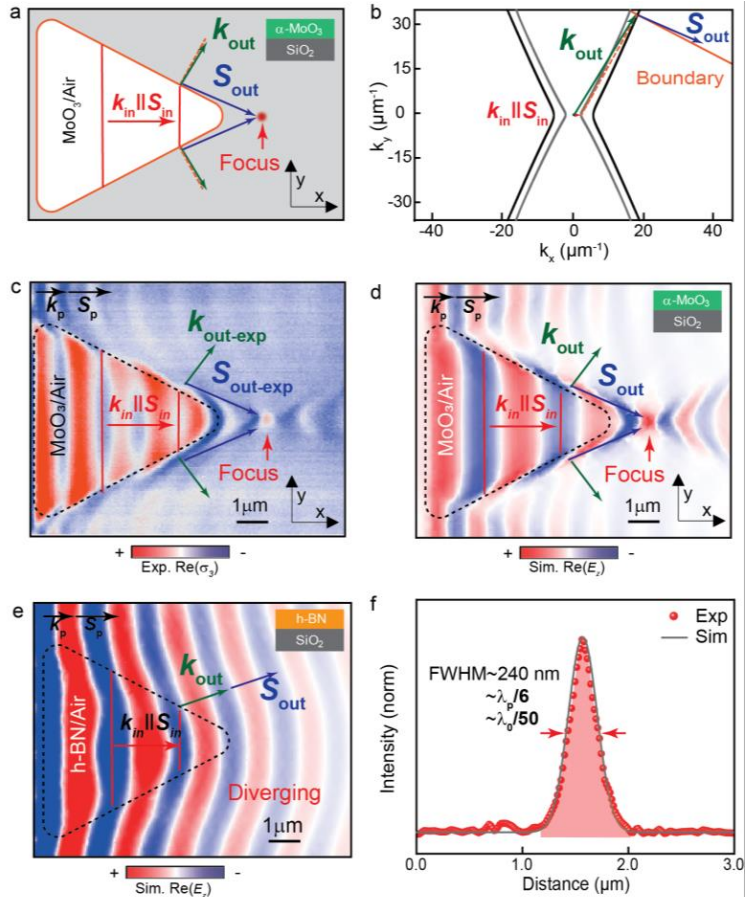

**Fig. 3 | Sub-diffractional planar lens based on refraction of HPhPs. a,** Schematics of a refractive hyperbolic lens fabricated by etching a triangular air cavity on the $SiO_2$ substrate below the α-$MoO_3$ flake. The top and bottom boundaries of the lens have the same slope as the boundary in (b). Upon refraction at the boundaries (orange contour line), polaritons bend far away from the normal, $S_{out}$ (blue arrows), with a tilted wavevector $k_{out}$ (green arrows), converging at a focal spot (red dot). **b,** Analytic IFCs of polaritons propagating in α-$MoO_3$/$SiO_2$ (black hyperbola) and α-$MoO_3$/air (grey hyperbola). When the boundary (orange line) is nearly perpendicular to the asymptote of the open hyperbolic IFC, the refracted polaritons propagate ($S_{out}$, blue arrows) almost parallel to the boundary with extremely large non-collinear wavevector ($k_{out}$, green arrows). **c,** Experimental near-field image of the the refractive planar hyperlens (black dashed line) for polaritons in a 170-nm-thick α-$MoO_3$ slab, at $\lambda_0 = 11.16$ μm. The polaritons converge upon refraction at the triangular boundary. Compared to non-refracted polaritons, indicated by $k_p$ and $S_p$ (black arrows) the refracted polaritons, $S_{out}$ (blue arrows), propagate nearly parallel to the boundary. **d,** Simulated near-field image of the refractive planar hyperlens (black dashed line) considered in (b) and visualized in (c). **e,** Simulated near-field image of a refractive lens for in-plane isotropic polaritons in a 170-nm-thick h-BN slab, at $\lambda_0 = 6.5$ μm**.** Upon refraction at the triangular boundary (black dashed line), polaritons bend towards the normal, $S_{out}$ (blue arrow**)**, with collinear wavevector $k_{out}$ (green arrow), yielding a diverging effect. Horizontal propagation of non-refracted polaritons is marked as $k_p$ and $S_p$. **f,**

Near-field intensity profiles (grey line and red dots) extracted through the focus spot along the vertical direction in (d) and (c), respectively. Both curves are normalized to the near-field intensity far away from the lens and flake edges. Confinement factors as large as $\sim\lambda_p/6$ and $\sim\lambda_0/50$ are obtained with respect to the polariton and free-space light wavelengths.

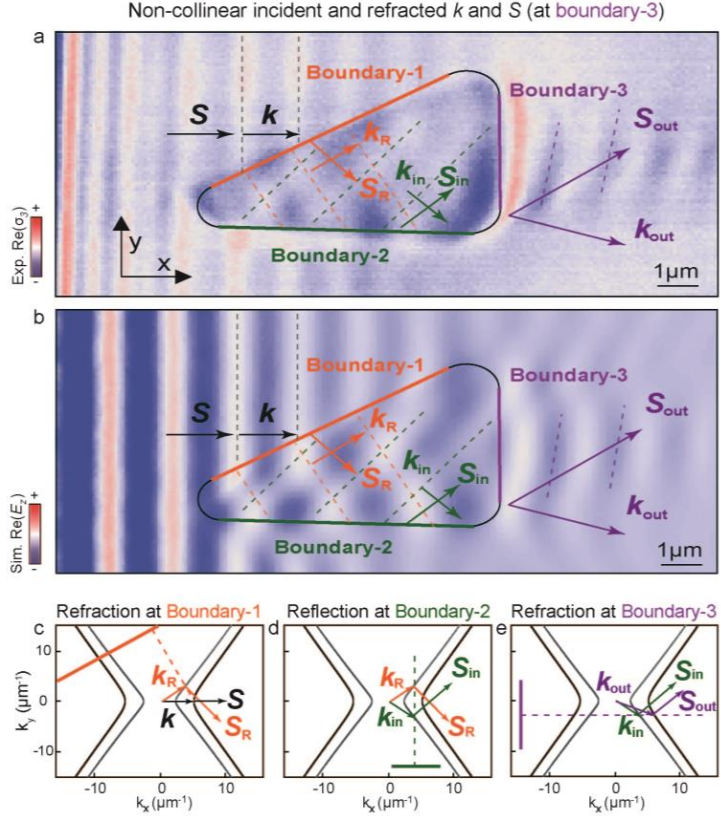

**Fig. 4 | Real-space visualization of the most general case of refraction between two anisotropic media using nanoscale-confined HPhPs passing through a bending-free planar prism. a,** Experimental $\text{Re}(\sigma_3(x,y))$ near-field images of polaritons propagating in a 231-nm-thick α-MoO$_3$ flake at $\lambda_0 = 11.0$ μm. The black contour line marks a triangular prism fabricated by etching an air cavity on the SiO$_2$ substrate below the α-MoO$_3$ flake. A first refraction takes place at boundary-1 (orange solid line) for incident polaritons with collinear **k** and **S** (black arrows), yielding refracted polaritons with non-collinear $\mathbf{k_R}$ and $\mathbf{S_R}$ (orange arrows). These polaritons then reflect at boundary-2 (green solid line), yielding polaritons with non-collinear $\mathbf{k_{in}}$ and $\mathbf{S_{in}}$ (green arrows). A second refraction at boundary-3 (violet solid line), yields polaritons with non-collinear $\mathbf{k_{out}}$ and $\mathbf{S_{out}}$ (violet arrows). **b,** Simulated $\text{Re}(E_z(x,y))$ near-field images of HPhPs for the case shown in (a). Dashed lines in experimental and simulated near-field images indicate the wavefronts of polaritons as they pass through the prisms. **c-e,** Analytic IFCs of HPhPs in MoO$_3$/air (grey curve) and MoO$_3$/SiO$_2$ (black curve), predicting the directions of refraction or reflection of HPhPs at boundary-1 (c), boundary-2 (d) and boundary-3 (e) based on momentum conservation.

So far, we have visualized refraction in hyperbolic media for the case in which collinear $\mathbf{k_{in}}$ and $\mathbf{S_{in}}$ refract anomalously into polaritons with non-collinear $\mathbf{k_{out}}$ and $\mathbf{S_{out}}$. However, the most general phenomenon of refraction involves the incident polaritons exhibiting non-collinear $\mathbf{k_{in}}$ and $\mathbf{S_{in}}$ (as sketched in Fig. 1b, and 1d). In the following, we study this fundamental phenomenon in hyperbolic media (Fig. 4). To do this, we again fabricate prisms in α-MoO$_3$ (following the same structure design as in Fig. 2) and visualize (by s-SNOM) the propagation of HPhPs refracting upon them. As shown in the near-field image of Fig. 4a, we observe that HPhPs launched by the edge of the flake (black arrows) refract at boundary-1 (Fig.4c), and the outcoming HPhPs propagate with non-collinear $\bm{k_R}$ and $\bm{S_R}$ inside the prism (orange arrows). As such, these polaritons can now be used to visualize the most general case of refraction at another boundary of the prism. However, to carry out this experiment successfully, we need to ensure that: *i*) the angle of the boundary allows refraction of polaritons according to momentum conservation, and *ii*) the HPhPs can reach this boundary within a reasonable propagation distance. We fulfill these conditions by considering a horizontal boundary (boundary-2) followed by a vertical boundary (boundary-3) in the triangular prism, as shown in the near-field image of Fig. 4a. In particular, we observe that HPhPs with non-collinear $\bm{k_R}$ and $\bm{S_R}$ are reflected at boundary-2 (Fig. 4d), yielding polaritons with non-collinear $\bm{k_{in}}$ and $\bm{S_{in}}$ (green arrows) propagating directly towards boundary-3, and reaching it within a reasonably short distance. Consequently, these polaritons refract at boundary-3, which results in polaritons with non-collinear $\bm{k_{out}}$ and $\bm{S_{out}}$ (violet arrows), as predicted by momentum conservation (Fig. 4e), and in good agreement with numerical simulations mimicking the experiment (Fig. 4b). This result unambiguously demonstrates refraction of waves whose energy flux and wavevector directions are non-collinear, and thus constitutes the first real-space visualization of the most general case of refraction between two hyperbolic media, which, in addition, we demonstrate at the nanoscale and in a natural medium. Furthermore, we highlight that after refraction at boundary-3, $\bm{S_{in}}$ and $\bm{S_{out}}$ are almost parallel, and waves refracted upon it behave as if they had been directly transmitted without any change in their direction of propagation. Note, however, that the wavevector does change upon this refraction phenomenon. Therefore, refraction upon this prism is bending-free, in excellent agreement with our theoretical prediction shown in Fig. 1d. This bending-free refractive prism opens exciting possibilities to engineer polaritonic wavefronts at the nanoscale without the need of changing their direction of propagation.

In summary, our work explores the anomalous character of refraction between strongly anisotropic media, which, despite its fundamental importance, has remained elusive up to now. Our observations of refraction of strongly-confined polaritons in a hyperbolic biaxial van der Waals crystal reveal an unprecedented optical effect: bending-free refraction, which pave the way for on-demand steering of light at the nanoscale and in natural media. Furthermore, our demonstration of a subwavelength planar lens at the nanoscale based on anomalous refraction constitutes the first member of a new class of planar nano-optical elements in anisotropic media. Altogether, our results open new avenues for integrated flat optics, directional energy transfer and heat management applications, as well as for mid-infrared (bio) sensing.